\documentclass[reprint,aps,superscriptaddress,prl]{revtex4-1}
\pdfoutput=1
\usepackage{graphicx,amsmath,amssymb,amscd,epsfig}
\usepackage{color}

\usepackage{placeins}
\usepackage[resetlabels]{multibib}
\newcites{main,suppl}{main,suppl}

\begin{document}
\title{Robust Type-II Weyl Semimetal Phase in Transition Metal Diphosphides XP$_2$ (X = Mo, W)}
\author{G. Aut\`es}
\affiliation{Institute of Physics, Ecole Polytechnique F\'ed\'erale de Lausanne (EPFL), CH-1015 Lausanne, Switzerland}
\affiliation{National Center for Computational Design and Discovery of Novel Materials MARVEL, Ecole Polytechnique F\'ed\'erale de Lausanne (EPFL), CH-1015 Lausanne, Switzerland}
\author{D. Gresch}
\affiliation{Theoretical Physics and Station Q Zurich, ETH Zurich, 8093 Zurich, Switzerland}
\author{M. Troyer}
\affiliation{Theoretical Physics and Station Q Zurich, ETH Zurich, 8093 Zurich, Switzerland}
\author{A. A. Soluyanov}
\affiliation{Theoretical Physics and Station Q Zurich, ETH Zurich, 8093 Zurich, Switzerland}
\affiliation{Department of Physics, St. Petersburg State University, St. Petersburg, 199034 Russia}
\author{O. V. Yazyev}
\affiliation{Institute of Physics, Ecole Polytechnique F\'ed\'erale de Lausanne (EPFL), CH-1015 Lausanne, Switzerland}
\affiliation{National Center for Computational Design and Discovery of Novel Materials MARVEL, Ecole Polytechnique F\'ed\'erale de Lausanne (EPFL), CH-1015 Lausanne, Switzerland}
\date{\today}
\begin{abstract} 
The recently discovered type-II Weyl points appear at the boundary between electron and hole pockets. Type-II Weyl semimetals that host such points are predicted to exhibit a new type of chiral anomaly and possess thermodynamic properties very different from their type-I counterparts.
In this Letter, we describe the prediction of a type-II Weyl semimetal phase in the transition metal diphosphides MoP$_2$ and WP$_2$. 
These materials are characterized by relatively simple 
band structures with four pairs of type-II Weyl points. Neighboring Weyl points have the same chirality, which makes the predicted topological
phase robust with respect to small perturbations of the crystalline lattice. In addition, this peculiar arrangement of the Weyl points
results in long topological Fermi arcs, thus making them readily accessible in angle-resolved photoemission spectroscopy.
\end{abstract}


\maketitle

Topological semimetals host band degeneracies in the vicinity of the Fermi level ($E_\mathrm{F}$) that are associated with certain integer-valued topological invariants~\citemain{volovik-book,wan11,bur11,wan12,you12,liu14,liu14a,taasprx,taasncomm}. Since these invariants cannot continuously change their values, the associated degeneracies are protected against perturbations. One such semimetal phase -- the Weyl semimetal (WSM) -- hosts point-like linear  band crossings of two bands with linear dispersion, the so-called Weyl points (WPs), in the vicinity of $E_\mathrm{F}$. These WPs represent sources or sinks of Berry curvature, and their associated topological invariant  is the Chern number $\mathcal{C}=\pm 1$ computed on a surface in momentum space that encloses the WP. Positive (negative)  Chern numbers correspond to a source (sink) of the Berry curvature. A variety of topology-driven physical phenomena is predicted and observed in WSMs, ranging from the observation of open Fermi arcs in the surface spectrum~\citemain{wan11,volovik-arcs} to the realization of the chiral anomaly of quantum field theory~\citemain{nie83,zyu12,hos13,vol14,xio15,zha15,hua15}.    

It was recently shown~\citemain{wte2} that unlike standard Lorentz-invariant field theory, condensed matter physics has two distinct types of Weyl fermions, and hence WSMs. While standard type-I Weyl fermions with closed Fermi surfaces were discovered in materials of the TaAs family~\citemain{taasprx,taasncomm,taassci,taasprx2,taasnat,taasnat2,nbas,tap,tap2}, the novel type-II Weyl fermions appear at the boundary between electron and hole pockets, leaving an open Fermi surface which results in the anisotropic chiral anomaly~\citemain{Volovik2014,wte2}. Two representatives of type-II WSMs materials considered to date are the orthorhombic low-temperature phases of WTe$_2$ and MoTe$_2$~\citemain{wte2,mote2,mote22,hasan-motewte}. Type-II Weyl points were also recently predicted to exist in strained HgTe~\cite{hgte}.
Eight (four) type-II WPs appear in WTe$_2$ (MoTe$_2$) formed by the valence and conduction bands. In WTe$_2$ some of the carrier pockets become topologically non-trivial, while in MoTe$_2$ they are all trivial, and the two materials represent very different Fermi arc arrangements~\citemain{wte2,mote22}. In both cases, however, the band structure is very complicated and the arrangement of WPs is sensitive to small changes in the crystal structure, which, in turn, is sensitive to temperature~\citemain{mote2, mote22}. Moreover, the proximity of WPs with opposite Chern numbers in $k$-space, as well as the existence of crossings between bands other than the topmost valence and the lowest conduction, at energies close to that of the WPs, makes the experimental confirmation of the type-II Weyl phase in WTe$_2$ and MoTe$_2$ a challenging task. The identification of materials with stable and easily observable type-II WPs thus sets 
an important problem in the study of this new topological phase.

In this Letter, we predict the existence of the type-II WSM phase in the previously synthesized compounds MoP$_2$ and WP$_2$. The crystal structure of these compounds is different from the previously reported ditellurides, resulting in a simpler band structure around $E_{\mathrm F}$ and a peculiar arrangement of WPs with nearest nodes characterized by the same Chern numbers, thus being stable against annihilating each other upon small lattice perturbations. This stability results in the presence of robust clearly visible long Fermi arcs at the surfaces of these compounds, which we expect to be readily observable in  angle-resolved photoemission spectroscopy (ARPES) experiments.

\begin{figure*}[hbtp]
\includegraphics[width=0.8\linewidth]{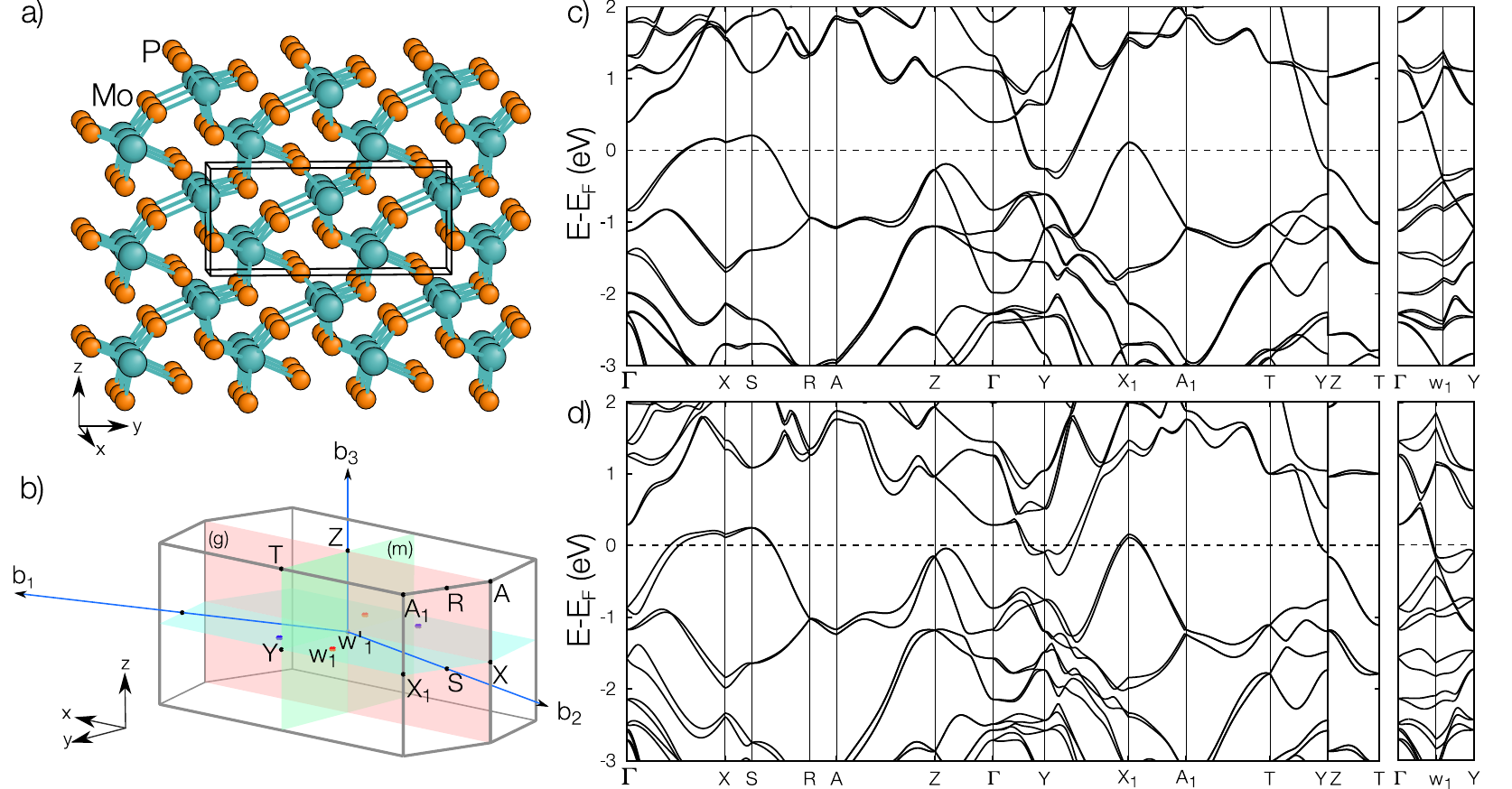}
\caption{(a) Crystal structure of MoP$_2$. The black box correspond to the orthorhombic conventional unit cell. (b) Brillouin zone of MoP$_2$ showing the positions of  Weyl nodes with positive (red) and negative (blue) Chern numbers.  Band structures of (c) MoP$_2$  and (d) WP$_2$  plotted along  high symmetry directions, as well as along the $\Gamma w_1 Y$ path in order to reveal the band crossings. }
\label{fig1}
\end{figure*}

The two compounds were identified by performing a high-throughput screening of the band structure topology of materials in the Inorganic Crystal Structure Database (ICSD)~\citemain{icsd}, using  the hybrid Wannier charge center technique~\citemain{sol11,yu11} as implemented in the Z2Pack package~\citemain{z2pack}. Both MoP$_2$ and WP$_2$ crystallize in an orthorhombic base-centered structure~\citemain{run63,rue83} containing two formula units per unit cell as shown in Fig.~\ref{fig1}a. Both crystals are non-centrosymmetric and belong to the non-symmorphic space group $Cmc2_1$ (36), which contains three symmetries:
a $C_2$ screw axis along the [001] direction ($2_1$), a mirror plane normal to the [100] direction ($m$, shown in green in Fig.~\ref{fig1}b), and a glide plane normal to the [010] direction ($g$, shown in red in Fig.~\ref{fig1}b). Although the crystalline symmetries of MoP$_2$ and WP$_2$ are similar to that of the ditellurides, the atomic structure of the phosphides is very different. While the structure of XTe$_2$ consist of van der Waals bonded layers, no layered structure exists in XP$_2$ materials. 

\begin{figure}[hbtp]
\includegraphics[width=\linewidth]{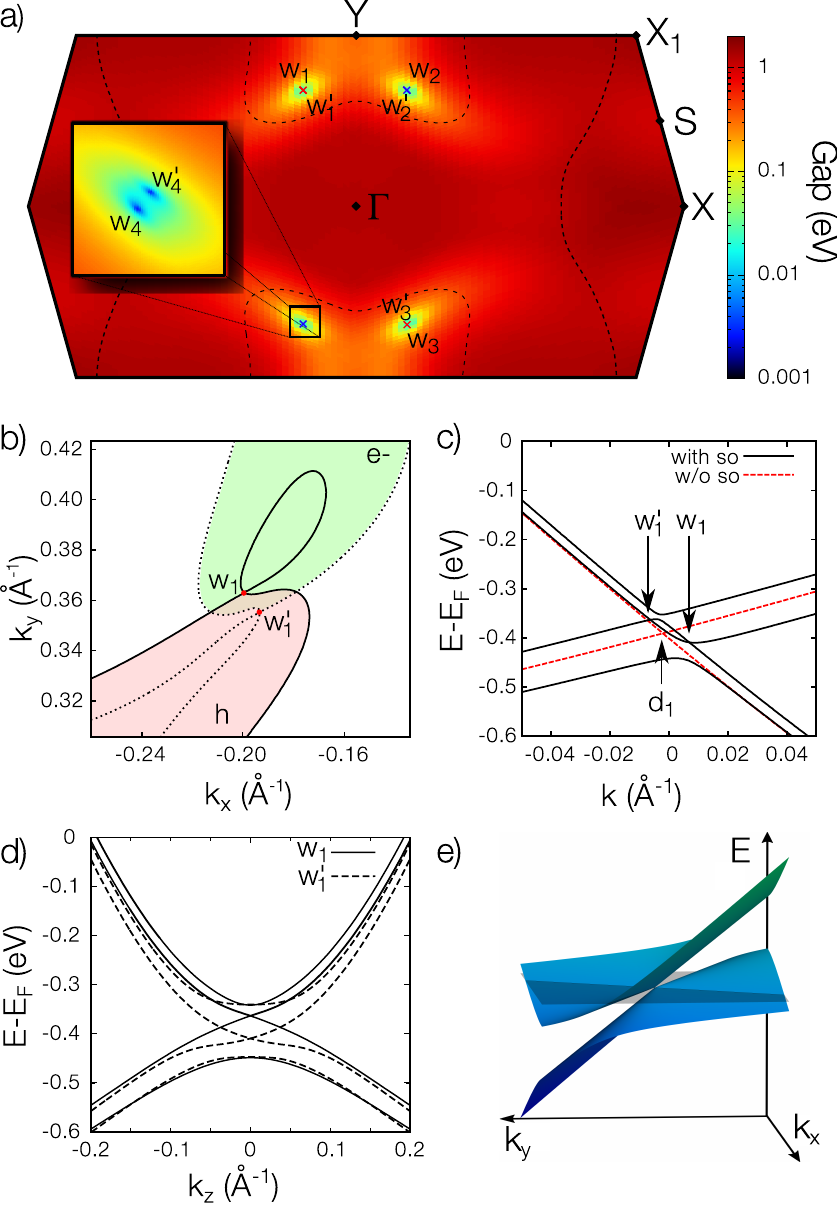}
\caption{(a) Energy difference between the lowest conduction band and the highest valence band of MoP$_2$ in the $k_z=0$ plane of the Brillouin zone.  The crosses correspond to the 8 Weyl nodes with Chern numbers $\mathcal{C}=+1$ (red) and $-1$ (blue). The dashed lines show the contours of the hole and electron pockets at the Fermi level in the $k_z=0$ plane. (b) Constant energy contour of the MoP$_2$ hole (red) and electron (green) pockets in the $k_z=0$ plane at the energy of $w_1$ (continuous line) and of $w_1'$ (dashed line). (c)  Band structure of MoP$_2$ along the $w_1w_1'$ line in the $k_z=0$ plane with (black) and without (red) spin-orbit coupling. (d) Band structure of MoP$_2$ along $k_z$ at $k_x$ and $k_y$ corresponding to the Weyl points $w_1$ (continuous line) and $w_1'$ (dotted line). (e) Energy dispersion around the type-II Weyl point $w_1$ in the $k_z=0$ plane.}
\label{fig2}
\end{figure}

The electronic structure of MoP$_2$ and WP$_2$ was computed from first principles~\citemain{qe,dal05}. The methodology  is described in more detail in the Supplemental Material~\footnote{See Supplemental Material at ... for the details of the first-principles calculations, the  derivation of the full model Hamiltonian, the computation of topological invariants and  a description of the additional Weyl points present in the valence band.}. The band structures along the high-symmetry directions of the Brillouin zone (BZ) are shown in Fig.~\ref{fig1}c and \ref{fig1}d, for MoP$_2$ and WP$_2$, respectively. Both compounds share a similar semimetallic band structure, very different from that of the XTe$_2$ compounds, with an electron pocket around $Y$ and a hole pocket in the vicinity of the $XS$ direction. The Fermi contour of these pockets is shown with a dashed line in Fig.~\ref{fig2}a for MoP$_2$ at $k_z=0$. The main difference between the two compounds is that the spin-orbit coupling (SOC) is stronger in WP$_2$, which results in a larger band splitting compared to MoP$_2$.

The band structure of MoP$_2$ and WP$_2$ along the high symmetry lines of the BZ suggests that these compounds are ordinary semimetals. However, a more careful analysis reveals the presence of eight points in the $k_z=0$ plane where conduction and valence bands touch. This can be seen in Fig.~\ref{fig2}a, where we plot the energy difference between the lowest conduction and the highest valence bands in the $k_z=0$ plane of the BZ. The gap closes at two inequivalent points $w_1$ and $w_1' $, located away from any high-symmetry line. The positions of these points are listed in Table~\ref{tab1}. The six other points $w_i$ and $w_i' $ ($i=2,3,4$) are related to $w_1$ and $w_1' $ by mirror and time-reversal ($\mathcal{T}$) symmetries.  In both compounds,   $w_1$ and $w_1' $ are at $-0.410$~eV and $-0.364$~eV relative to the Fermi level in MoP$_2$  and at $-0.471$~eV and $-0.340$~eV relative to the Fermi level in WP$_2$.

\begin{table}[b]
\caption{Positions, Chern numbers and energies of the Weyl points in MoP$_2$ and WP$_2$.} 
\begin{tabular}{llcccc}
\hline\hline  & & $k_x$  & k$_y$& Chern number& $E-E_{\textrm F}$    \\ 
              &  & (\AA$^{-1}$)& (\AA$^{-1}$) & $\mathcal{C}$ & (eV) \\ 
\hline 

 MoP$_2$ & $w_1$    & $-0.2010$ &  $0.3627$  & $+1$ & $-0.410$ \\ 
 MoP$_2$ & $w_{1}' $ & $-0.1939$ &  $0.3516$  & $+1$ & $-0.364$ \\ 
\hline
  WP$_2$  &$w_1$  & $-0.2627$  & $0.3165$   & $+1$ & $-0.471$ \\ 
 WP$_2$ & $w_{1}'$ & $-0.2577$  & $0.2818$   & $+1$ & $-0.340$ \\ 
\hline\hline
\end{tabular}

\label{tab1} 
\end{table}

\begin{figure*}[ht]
\includegraphics[width=1.0\linewidth]{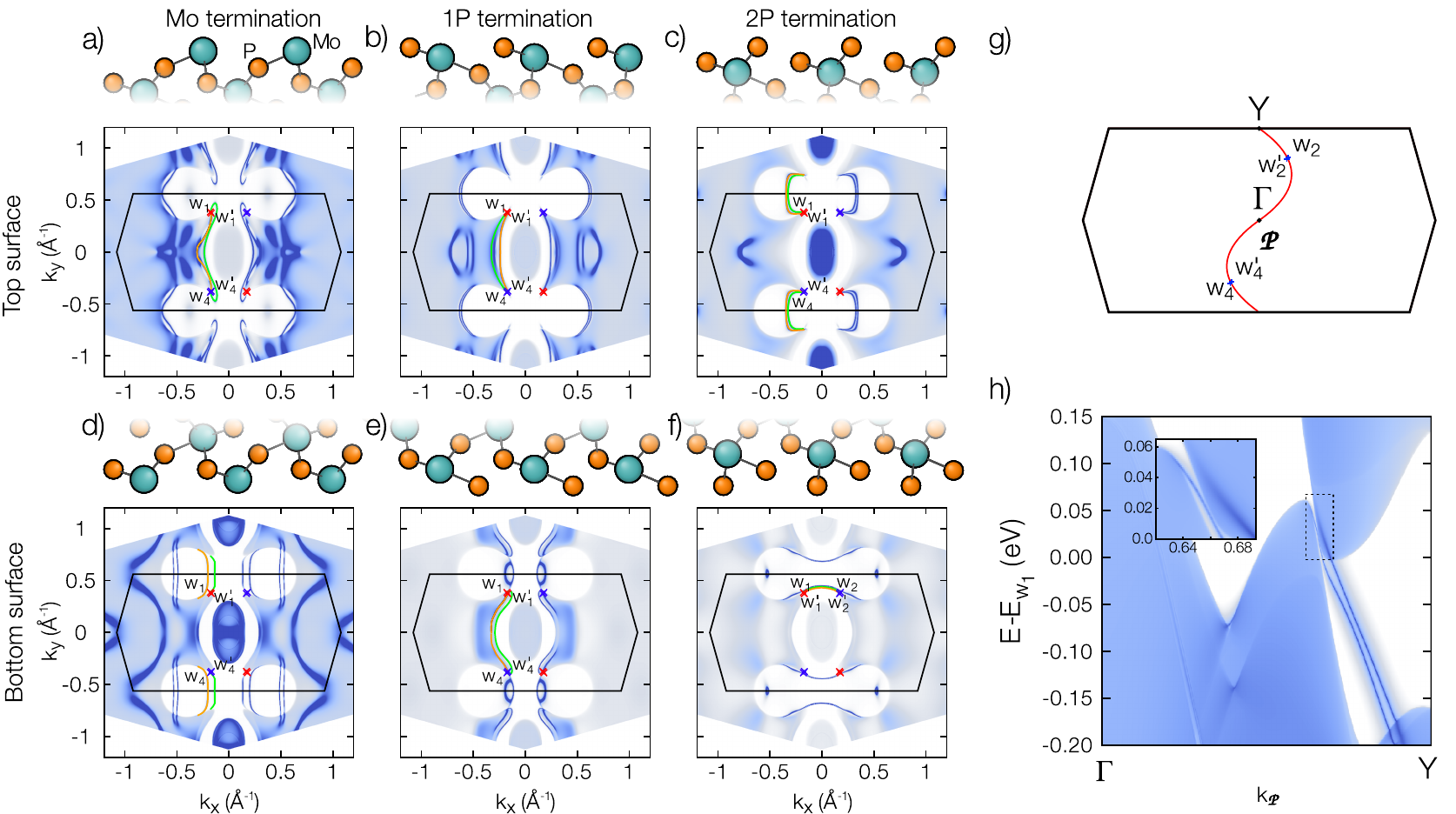}
\caption{ Surface density of states of (a,b,c) the top and (d,e,f) the bottom (001) surfaces with Mo- and P-terminations 1P and 2P  at the energy of the Weyl node $w_1$. The green and orange lines indicate the topological Fermi arcs connecting the $w_i$ and $w_i'$ WPs, respectively. (g) Cut in the $k_z=0$ plane of the $\mathcal{T}$-symmetric plane ${\cal P}$ with $\mathbb{Z}_2$ invariant 1. (h) Surface density of states of the top P-terminated surface 2P along $\mathcal{P}$ from $\Gamma$ to $Y$. The inset panel shows a magnified view of the region in proximity of the hole and electron pockets. Only one of the two surface states is topological, connecting the electron and hole pockets.}
\label{fig3}
\end{figure*}

Analogously to the case of XTe$_2$, the existence of degeneracy points in the $k_z=0$ plane of MoP$_2$ and WP$_2$ is due to the presence of the product symmetry $C_2 \mathcal{T}$, which restricts a general $2 \times 2$ Hamiltonian in the plane to be of the form
\begin{equation}
\mathcal{H}(k_x, k_y, 0) = d_0(k_x,k_y) \sigma_0 + d_y(k_x,k_y) \sigma_y + d_z(k_x,k_y) \sigma_z,
\label{ham}
\end{equation}
where $\sigma_{y,z}$ are the corresponding Pauli matrices and $\sigma_0$ is the 2$\times$2 unit matrix associated with the kinetic energy term of the type-II Weyl Hamiltonian~\citemain{wte2}. The full derivation
of the model Hamiltonian is presented in the Supplemental Material~\citemain{Note1}.
In order to establish that the degeneracies $w_i$ and $w_i'$ are indeed WPs, we computed the Chern numbers of surfaces enclosing these points following the method described in Ref.~\citemain{wte2}. We find that both $w_1$ and $w_1'$ carry a topological charge  $\mathcal{C}=+1 $~\citemain{Note1}, while the charges of the other six points are obtained by symmetry arguments: mirror reflection flips the sign of the Chern number of a WP, thus $w_{2,4}$ and $w_{2,4}'$ have $\mathcal{C}=-1$, while $\mathcal{T}$-reflection preserves it, so $\mathcal{C}=+1$ for $w_3$ and $w_3'$. 
 
The WPs in  MoP$_2$ and WP$_2$ are of type-II as can be concluded by examining the Fermi surface at the energies of $w_1$ and $w_1'$, shown in Fig.~\ref{fig2}b for the case of MoP$_2$. Both nodes appear at the points of contact between the electron pocket located around $Y$ and the hole pocket located along the $XS$ direction. To further confirm this conclusion, we fitted the {\it ab initio} band structure in the vicinity of the WPs to find the coefficients $d_i$ of Eq.~\ref{ham} to the linear order in ${\bf k}$~\citemain{Note1}. For both the $w_1$ and $w_1'$ points, the kinetic term dominates the spectrum along the $k_y$ direction around the WP, as illustrated in Fig.~\ref{fig2}e for the WP $w_1$. 
The dominant kinetic term in the $k_y$ direction suggests a possible observation of the type-II chiral anomaly~\citemain{wte2} in XP$_2$ compounds when both electric and magnetic fields are applied along this direction.

The WPs $w_1$ and $w_1'$ are separated in energy. To understand the origin of this separation, we note that the two points are formed by spin-split bands, as can be concluded from Fig.~\ref{fig2}c and Fig.~\ref{fig2}d, where the energy dispersion of MoP$_2$ is shown along a path connecting the two WPs in the $k_z=0$ plane and along the $k_z$ direction, respectively. Indeed, for a calculation performed without taking into account SOC, the electron and hole pockets touch at 4 crossing points $d_i$ in the $k_z=0$ plane, which have the associated topological charge $\mathcal{C}=\pm 2$ and correspond to the superposition of two WPs of the same chirality, as expected for $\mathrm{SU}$(2) symmetry~\citemain{route}. These double WPs are split by the SOC into single nodes $w_i$ and $w_i'$ (see Fig.~\ref{fig2}c) that have the same chirality.

The magnitude of the splitting in energy and $k$-space between the two adjacent WPs $w_i$ and $w_i'$ is hence directly related to the strength of the SOC. These splittings are larger in WP$_2$ (131~meV and 0.035~\AA$^{-1}$) than in MoP$_2$ (46~meV and 0.013~\AA$^{-1}$) (see Table~\ref{tab1}), as expected due to the larger SOC strength in W. This suggests the possibility of tuning the separation between the WPs in these compounds in both energy and momentum by applying strain or chemical substitution, since this changes the effective SOC.

Unlike the case of ditellurides, the neighboring WPs in XP$_2$ materials have the same chirality, and thus cannot annihilate each other. This implies that MoP$_2$ and WP$_2$ realize a stable  type-II WSM phase that is far from a possible topological phase transition caused by a merging of the opposite chirality WPs. Opposite chirality WPs  can annihilate when they reach the same point of the 3D BZ, so the smallest distance in $k$-space between WPs with opposite Chern number $\mathcal{C}$ can be considered as a measure of stability of the WSM phase. We find this distance to be 0.38\AA$^{-1}$  and 0.52\AA$^{-1}$ in MoP$_2$ and WP$_2$, respectively, which constitutes $20$\% and $26$\% of the corresponding inverse lattice constants. These numbers can be compared to the distance between opposite chirality WPs in the TaAs  materials family, in particular TaP where the distance is the longest\citemain{taasprx} and   is  0.09\AA$^{-1}$ ($4$\% of the inverse lattice constant). The distance between the neighboring opposite chirality WPs in XTe$_2$ is $0.7$\% of the inverse lattice constant.   

One evident consequence of the large $k$-space separation of opposite chirality WPs in XP$_2$ is the possibility of the observation of extended topological Fermi arcs in the surface  of these materials. For type-I WPs, which have a point-like Fermi surface, these Fermi arcs connect the projections of the opposite chirality WPs onto the surface. In contrast,	 the Fermi surface of the type-II WPs is open and the projection of the points is generally hidden within the projection of a charge-carrier pocket.     

We consider the (001) surface of MoP$_2$ and WP$_2$, since the eight WPs project onto distinct points of the corresponding surface BZ, allowing for the observation  of Fermi arcs. We find that at all energies all electron and hole pockets enclose an equal number of chiral and antichiral WPs, therefore the Chern number of all these pockets vanishes. From this perspective, no Fermi arcs connecting electron and hole pockets are expected. Indeed, at $E_{\mathrm{F}}$ we find that the projected WPs are covered by the projection of the electron pocket around the $Y$ point and no Fermi arcs can be observed. However, we do find topological Fermi arcs in these materials at lower energies as we explain below.  
 
We computed the surface density of states of MoP$_2$ using the tight-binding model, obtained from the bulk Wannier functions~\citemain{wannier} for Mo $d$  and P $p$ orbitals, in order to compute the Green's function of the semi-inifnite surface according to the method introduced in Ref.~\citemain{umerski}.  Three possible surface terminations were investigated: one Mo-terminated and two P-terminated denoted as 1P and 2P. Furthermore, the top and bottom (001) surfaces of MoP$_2$ are inequivalent, thus giving rise to six different configurations (see Fig.~\ref{fig3}a--f). The surface densities of states at the energy of $w_1$, $0.41$~eV below $E_{\mathrm{F}}$, all show two distinct Fermi arcs. One of these arcs connects two $w_i$  points of opposite chirality (green line in Fig.~\ref{fig3}a--f), while the other one connects  two $w_i'$  points of opposite chirality (orange line in Fig.~\ref{fig3}a--f).
Both the connectivity and the shape of the Fermi arcs depend on the details of the surface termination. For the Mo and 1P top surfaces as well as for the 1P bottom surface, the Fermi arcs connect $w_1$ ($w_1'$) and $w_4$ ($w_4'$) within the surface BZ, which is shown as black hexagon in Fig.~\ref{fig3}. On the 2P top surface and the Mo bottom surface  $w_1$ ($w_1'$) is connected by a Fermi arc to $w_4$ ($w_4'$) across the surface BZ boundary. Finally, on the 2P bottom surface, the Fermi arcs connect the $w_1$ ($w_1'$) and $w_2$ ($w_2'$) WPs.

While the connectivity of these Fermi arcs depends on a particular surface, one can argue that the arcs themselves are of topological origin. In order to demonstrate this we computed the $\mathbb{Z}_2$ invariant on the $\mathcal{T}$-symmetric plane ${\cal P}$ shown in Fig.~\ref{fig3}g and find it to be non-trivial (see Supplemental Material for details~\citemain{Note1}). This implies that this cut of the BZ, on which the spectrum is gapped at all points, is a quantum spin Hall effect system~\citemain{kan05}, and therefore it is guaranteed to show topologically protected edge states. In other words, any line cut of the (001) surface BZ that passes between the two WPs  has to cross a topological Fermi arc as shown in Fig.~\ref{fig3}h. The conclusion that the Fermi arcs connecting points $w_i$ and $w_i'$ are of topological origin agrees with the values of $\mathbb{Z}_2$ invariants defined on different $\mathcal{T}$-symmetric planes of the BZ, as further discussed in the Supplemental Material~\citemain{Note1}.  

The predicted Fermi arcs span extended regions of $k$-space and are localized below the Fermi level. Therefore they should be readily observable in ARPES experiments on a cleaved (001) surface of XP$_2$ compounds. Experimental observation of these arcs will provide an unambiguous evidence of the type-II Weyl fermions in these materials. It should be noted that other surface states are visible in Figs.~\ref{fig3}a--f. A careful examination of the electronic structure shows that these states do not originate from Weyl points and are thus topologically trivial, as discussed in the Supplemental Material~\citemain{Note1}.

In conclusion, we theoretically identified  a new family of type-II Weyl semimetals in the transition disphophides MoP$_2$ and WP$_2$. The Brillouin zone of these materials contains 4 pairs of Weyl nodes with the same chirality in the $k_z=0$ plane, which implies robustness of the predicted Weyl semimetal phase. 
We predict that a type-II chiral anomaly should be observable in these compounds and that long Fermi arcs should be detectable by ARPES experiments on  the [001] surface  with a great variety of possible arrangements depending on the surface termination.

{\it Acknowledgments} G.A. and O.V.Y. acknowledge support by the NCCR Marvel and the ERC Starting grant ``TopoMat'' (Grant No. 306504). First-principles electronic structure calculations have been performed at the Swiss National Supercomputing Centre (CSCS) under project s515. D. G., A. A. S. and M. T.  were supported by Microsoft Research, the European Research Council through ERC Advanced Grant SIMCOFE, and the Swiss National Science Foundation through the National Competence Centers in Research MARVEL and QSIT.

\bibliographystylemain{apsrev4-1}
\bibliographymain{main}

\clearpage

%
\setcounter{equation}{0}
\setcounter{figure}{0}
\setcounter{table}{0}
\setcounter{page}{1}
\makeatletter
%
\renewcommand{\theequation}{S\arabic{equation}}
\renewcommand{\thefigure}{S\arabic{figure}}
\renewcommand{\thetable}{S\arabic{table}}
\renewcommand{\bibnumfmt}[1]{[S#1]}
\renewcommand{\citenumfont}[1]{S#1}
%

%
\widetext
%
\begin{center}
\textbf{\large{Supplemental Material for\\ ``Robust Type-II Weyl Semimetal Phase in Transition Metal Diphosphides XP$_2$ (X = Mo, W)''}}
\end{center}
\section{I. Methodology}

The electronic structure of MoP$_2$ and WP$_2$ was computed within the density functional theory (DFT) framework using the generalized gradient approximation (GGA) as implemented in QUANTUM-ESPRESSO software package~\citesuppl{qe1}. Spin-orbit coupling (SOC) is taken into account with the help of  fully relativistic ultrasoft pseudopotentials~\citesuppl{dal051}. 
The calculations were carried out using an 8×8×5 k-point mesh and a planewave kinetic energy cutoff of 50 Ry for the wavefunctions. We used the experimentally determined crystal structure from Ref.~\citesuppl{rue831}.

\section{II. Model Hamiltonian of the Weyl points}

The general $2 \times 2$ Hamiltonian restricted by $C_2\mathcal{T}$ symmetry in the $k_z=0$ plane  is of the form~\citesuppl{wte21, route1}
\begin{equation}
\mathcal{H}(k_x, k_y, 0) = d_0 \sigma_0 + d_y \sigma_y + d_z \sigma_z.
\end{equation}
The co-dimension of the system of equations defining a band gap closure ($d_y = d_z = 0$) is zero, meaning that stable nodal points can exist.
The general linearized form of the Hamiltonian around a crossing point in the $k_z = 0$ plane is given by
\begin{equation}
\mathcal{H} = v_1 k_x + v_2 k_y + (a k_x + b k_y)\sigma_y + (c k_x + d k_y) \sigma_z + e k_z \sigma_x + \varepsilon_0,
\label{eq:H}
\end{equation}
where the $k_i$ are relative to the crossing point. 
The fitting parameters for the different Weyl points are given in Table \ref{tab:weyl_fit}. From these parameters, one can easily verify that all Weyl fermions are of type-II, as shown in Fig.~2e of the main text.

\begin{table}[h]
\caption{Linear fit parameters of the Hamiltonian around the Weyl nodes in $\text{eV}~\text{\AA}^{-1}$.}
\begin{tabular}{ccccccc}
\hline\hline
 & $b$ & $c$ & $d$ & $e$ & $v_1$ & $v_2$  \\\hline
MoP$_2$ $w_1$ & $1.11$ & $1.46$ & $-1.82$ & $-0.90$ & $0.44$ & $-2.6$ \\
MoP$_2$ $w_1'$ & $1.18$ & $1.43$ & $-1.32$ & $-0.77$ & $0.73$ & $-2.99$ \\\hline
WP$_2$ $w_1$ & $1.20$ & $1.79$ & $-1.29$ & $-0.06$ & $-0.47$ & $-3.49$ \\
WP$_2$ $w_1'$ & $1.94$ & $1.64$ & $-2.16$ & $-0.12$ & $-0.91$ & $-2.30$\\\hline\hline

\end{tabular}

\label{tab:weyl_fit}
\end{table}

\section{III. Chirality of the Weyl nodes and topological invariants}

\begin{figure}[hbtp]
\includegraphics[width=0.5\linewidth]{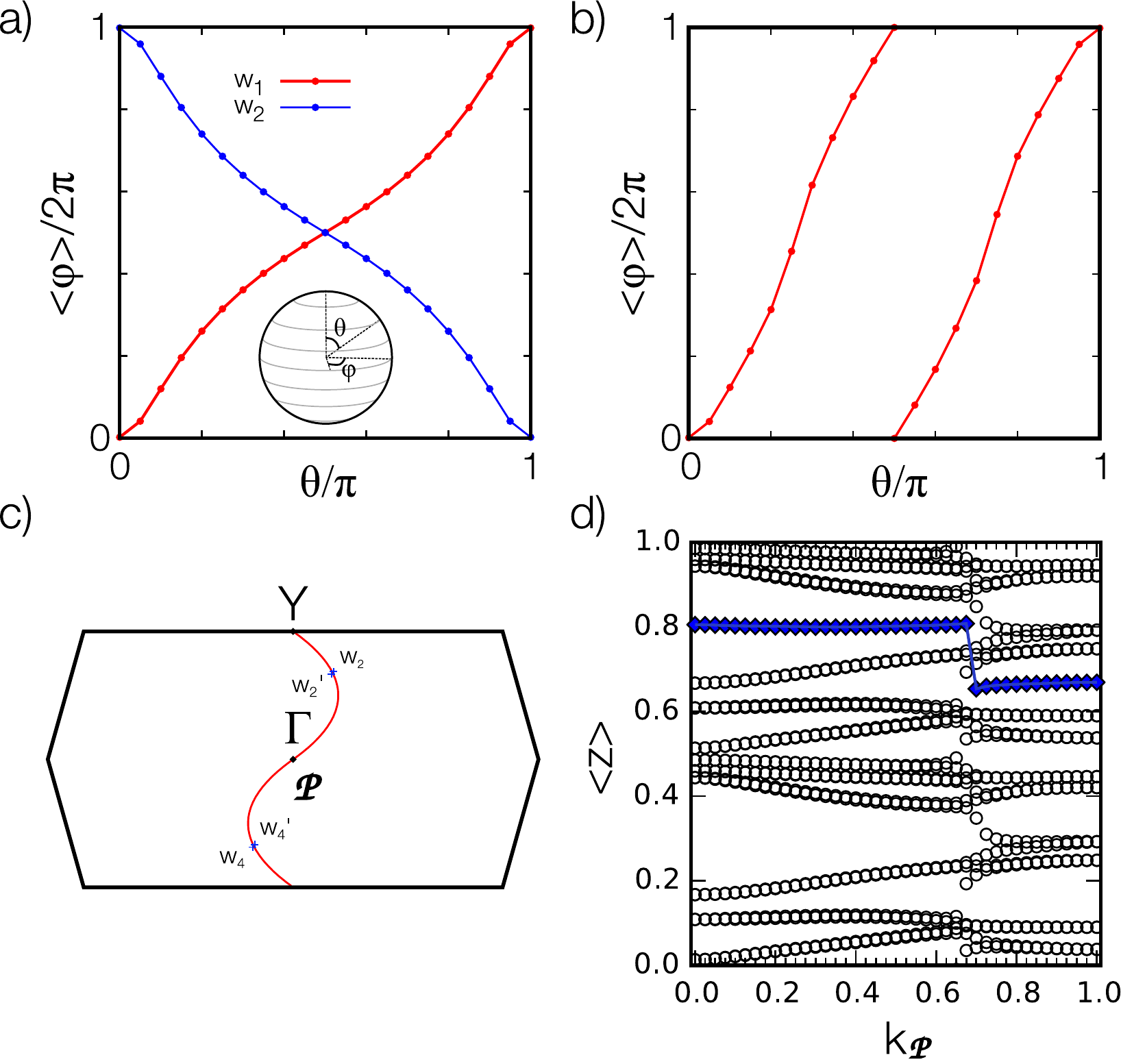}
\caption{ a) Evolution of the Wannier charge centers on loops of longitudinal angle $\theta$ forming a sphere enclosing $w_1$ (red) and a sphere enclosing $w_2$ (blue) in MoP$_2$, b) Evolution of the Wannier charge centers on loops of longitudinal angle $\theta$ forming a sphere enclosing $w_1$ and $w_1'$ (red)  in MoP$_2$, d) Wannier charge
centers $\langle z\rangle$ (black circle) along the half-plane  $(\mathcal{C},k_z)$ from $\Gamma$ to $Y$ shown in red in panel (c). The blue diamonds show the position of the largest gap which corresponds to a non-trivial $\mathbb{Z}_2$ invariant.}
\label{figsupp1}
\end{figure}

The chirality of the Weyl points $w_i$ and $w_i'$ was obtained by computing the flux of Berry curvature through a surface enclosing them.
In order to carry out this calculation, we follow the method proposed by Ref.~\citesuppl{wte21}. We calculate Wannier charge centers on longitudinal loops around a sphere enclosing the Weyl point (see Fig.~\ref{figsupp1}a). The sum of the Wannier charge centers on a loop with longitudinal angle $\theta_{i}$ corresponds to the Berry phase accumulated along the loop, or similarly to the average position  $\langle\phi\rangle$ of the charge on the loop. When the angle $\theta$ varies from 0 to $\pi$, the loops cover a closed surface and the average position of center  $\langle\phi\rangle$ can only be shifted by an integer number of $2\pi$. This number correspond to the chirality of the Weyl node enclosed in the sphere. 

 We apply this procedure to a sphere enclosing $w_1$ (red curve, Fig.~\ref{figsupp1}a).
 As the angle $\theta$ varies from 0 to $\pi$, the average position of the Wannier centers on the loops ($\langle\phi\rangle$) is shifted by $2\pi$,   which indicates  that $w_1$ is a source of Berry curvature with chirality $\mathcal{C}=+1$.
 The Weyl node $w_2$, which is the mirror image of $w_1$, is then expected to carry a topological charge $\mathcal{C}=-1$. 
 This is confirmed by the evolution of the WCC on a sphere enclosing $w_2$  (blue curve, Fig.~\ref{figsupp1}a) 
 The same method can be applied to show that the Weyl node $w_1'$ has a chirality $\mathcal{C}=+1$ (see Fig.~\ref{figsupp1}b).

To further elucidate the topological character of MoP$_2$ and WP$_2$, several topological invariants were computed from first-principles. 
The presence of time-reversal ($\mathcal{T}$) symmetry  allows the computation of the $\mathbb{Z}_2$ invariant on $\mathcal{T}$-symmetric planes in the BZ,
defined by $k_i = 0, \pi$, where $i = x, y, z$. The $k_z = 0$  hosts WPs and does not
have a well-defined invariant. For the other five planes, the topological invariant is computed using a hybrid Wannier centers technique~\citesuppl{sol111, yu111}
and found to be trivial. Nevertheless, it is still possible to define a non-trivial $\mathbb{Z}_2$ invariant~\citesuppl{kan051} by considering a curved $\mathcal{T}$-symmetric plane crossing between the $w_1$ and $w_1'$ (or $w_2$ and $w_2'$ ) and their $\mathcal{T}$ image  $w_3$ and $w_3'$ (or $w_4$ and $w_4'$). The path  along which such a plane $\mathcal{P}$ cut the $k_z=0$  plane is shown on Fig.~\ref{figsupp1}c. The evolution of the Wannier charge centers positions on the half plane $\mathcal{P}$ from $\Gamma$ to $Y$ is shown on Fig.~\ref{figsupp1}d which is not gapped and  corresponds to a non-trivial $\mathbb{Z}_2$ invariant. 

\section{IV. Additional topological features below the valence band}

An examination of  the band structure plots along the $\Gamma Y$ and $Y X_1$ direction in Fig.1 of the main text, reveals that other non-trivial crossings exist  between the bands below the valence and above the conduction band. In order to allow an easy experimental observation of the main WPs, it is important that such features are located sufficiently far in energy or $k$-space from the $w_i$ and $w_i'$ points. In the valence bands, 12 WPs corresponding to crossings between the $n-1$ and $n-2$ bands (where $n$ is the number of valence electrons) are located in the $k_z=0$ plane on each side of the $\Gamma Y$ line at $\approx 0.26$~eV below the $w_i$ WP. In the conduction band,  8 WPs   corresponding to crossings between the $n+2$ and $n+3$ bands are located in the $k_z=0$ plane at $\approx 0.8$~eV above the $w_i$ WP. The energies and position of these features show that the main WPs $w_i$ and $w_i'$ are well isolated from any others WPs. This provides a great advantage for the experimental characterization of the Weyl phase over other type-II WSM such as WTe$_2$ and MoTe$_2$.

\begin{figure}[hbtp]
\includegraphics[width=0.5\linewidth]{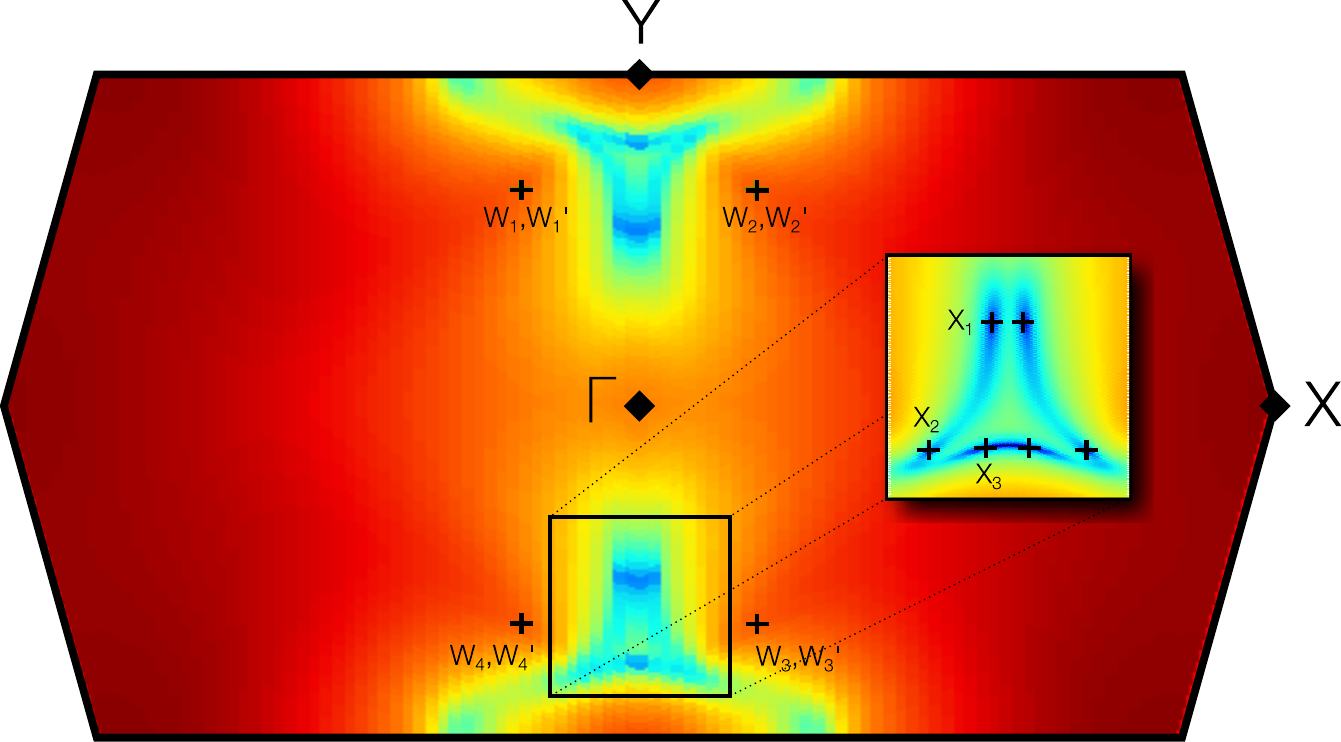}
\caption{ Energy difference between the $n-1$ and $n-2$ bands in the $k_z=0$ plane of MoP$_2$. }
\label{figsupp2}
\end{figure}

Here, we describe  additional topological features that appear below the valence band in MoP$_2$ and WP$_2$. The closest in energy to the Fermi level corresponds to crossing between the $n-1$ and $n-2$ bands (where $n$ is the number of valence electron). 12 Weyl points formed by these bands are located in the $k_z=0$ plane on each side of the $\Gamma Y$ line. The position of 3 inequivalent points $x_1$, $x_2$ and $x_3$ is shown on Fig.~\ref{figsupp2} for MoP$_2$. The 9 other points can be obtained by applying the mirror and time-reversal symmetries. 
The highest in energy of these points  has an energy of $-0.673$ eV  with respect to the Fermi level (see Table~\ref{tabsupp1}). They are thus separated from the lowest Weyl point $w_i$ by an energy of $\approx 0.26$ eV. The position in $k$-space and energy of the $x_i$ points demonstrate that the main Weyl points $w_i$ and $w_i'$ are well isolated and that the additional surface states that appears in Fig. 3a--f of the main text are not of topological origin. 

\begin{table}[th]
\caption{Positions and energies of the Weyl nodes between the $n-1$ and $n-2$ bands in MoP$_2$} 
\begin{tabular}{cccc}
\hline\hline  & $k_x$ (\AA$^{-1}$) & k$_y$ (\AA$^{-1}$) & $E-E_F$ (eV)   \\ 
\hline 

 $x_1$    & $-0.0192$ &  $-0.2933$  & $-0.684$ \\ 
 $x_2 $ & $-0.0923$ &  $-0.4465$  &  $-0.692$ \\ 
 $x_3$  & $-0.0232$  & $-0.4401$   &  $-0.673$ \\ 

\hline\hline
\end{tabular}

\label{tabsupp1} 
\end{table}

\bibliographysuppl{suppl}

\end{document}